\documentclass[fleqn,usenatbib]{mnras}
\usepackage{newtxtext,newtxmath}
\usepackage[T1]{fontenc}
\usepackage{ae,aecompl}

\usepackage{xcolor}
\usepackage{ragged2e}

\usepackage{subfig,graphicx,caption}
\usepackage{amsmath}	    

\title[Flashlights: An Off-Caustic Lensed Star]
{Flashlights: An Off-Caustic Lensed Star at Redshift \lowercase{z}=1.26 in Abell 370}

\author[A. K. Meena et al.]{
Ashish Kumar Meena$^{1}$\thanks{E-mail: \href{mailto:ashishmeena766@gmail.com}{ashishmeena766@gmail.com}},
Wenlei Chen$^{2}$, 
Adi Zitrin$^{1}$,
Patrick L. Kelly$^{2}$,
Miriam Golubchik$^{1}$,
Rui Zhou$^{2}$,
\newauthor
Amruth Alfred$^{3}$,
Tom Broadhurst$^{4,5,6}$,
Jose M. Diego$^{7}$,
Alexei V. Filippenko$^{8}$,
Sung Kei Li$^{3}$,
\newauthor
Masamune Oguri$^{9,10}$,
Nathan Smith$^{11}$,
and
Liliya L. R. Williams$^{2}$
\\
\\
$^{1}$Physics Department, Ben-Gurion University of the Negev, P.O. Box 653,Be'er-Sheva 8410501, Israel \\
$^{2}$School of Physics and Astronomy, University of Minnesota, 116 Church Street SE, Minneapolis, MN 55455, USA \\
$^{3}$Department of Physics, The University of Hong Kong, Pokfulam Road, Hong Kong \\
$^{4}$Department of Theoretical Physics, University of the Basque Country UPV/EHU, Bilbao, Spain \\
$^{5}$Donostia International Physics Center (DIPC), 20018 Donostia, Spain \\
$^{6}$IKERBASQUE, Basque Foundation for Science, Bilbao, Spain \\
$^{7}$Instituto de Física de Cantabria (CSIC-UC). Avda. Los Castros s/n. 39005 Santander, Spain \\
$^{8}$Department of Astronomy, University of California, Berkeley, CA 94720-3411, USA \\
$^{9}$Center for Frontier Science, Chiba University, 1-33 Yayoi-cho, Inage-ku, Chiba 263-8522, Japan \\
$^{10}$Department of Physics, Graduate School of Science, Chiba University, 1-33 Yayoi-Cho, Inage-Ku, Chiba 263-8522, Japan \\
$^{11}$Department of Astronomy, University of Arizona, Tucson, AZ 85721, USA
}

\pubyear{2022}

\begin{document}
\label{firstpage}
\pagerange{\pageref{firstpage}--\pageref{lastpage}}
\maketitle

\begin{abstract}
We report the discovery of a transient seen in a strongly lensed arc at redshift $z_{\rm s}=1.2567$ in \emph{Hubble Space Telescope} imaging of the Abell 370 galaxy cluster. The transient is detected at $29.51\pm0.14$ AB mag in a WFC3/UVIS F200LP difference image made using observations from two different epochs, obtained in the framework of the \emph{Flashlights} program, and is also visible in the F350LP band~($m_{\rm F350LP} \approx 30.53\pm0.76$~AB mag). The transient is observed on the negative-parity side of the critical curve at a distance of $\sim 0\farcs6$ from it, greater than previous examples of lensed stars. The large distance from the critical curve yields a significantly smaller macromagnification, but our simulations show that bright, O/B-type supergiants can reach sufficiently high magnifications to be seen at the observed position and magnitude. In addition, the observed transient image is a trailing image with an observer-frame time delay of $\sim+0.8$~days from its expected counterpart, so that any transient lasting for longer than that should have also been seen on the minima side and is thus excluded. This, together with the blue colour we measure for the transient ($m_{\rm F200LP} - m_{\rm F350LP} \approx [-0.3,-1.6]$~AB), rules out most other transient candidates such as (kilo)novae, for example, and makes a lensed star the prime candidate. Assuming the transient is indeed a lensed star as suggested, many more such events should be detected in the near future in cluster surveys with the \emph{Hubble Space Telescope} and \emph{James Webb Space Telescope}.
\end{abstract}


\section{Introduction}
\label{sec:Introduction}

Galaxy clusters are the most massive gravitationally bound structures in the Universe with high central mass densities that commonly generate strong gravitational lensing effects. High magnification of distant sources due to strong lensing has allowed the distant Universe to be probed, revealing a large number of high-redshift galaxies \citep[e.g.,][]{2013ApJ...762...32C, 2017ApJ...835...44C, 2018ApJ...864L..22S, 2022arXiv221014123Y,Roberts-Borsani2022z98, Williams2022z9p5, 2023MNRAS.518.4755A, 2023MNRAS.519.1201A}.

The \emph{critical curves} and \emph{caustics} of a gravitational lens are infinitely thin contours where the lens magnification diverges (for an infinitely small source) in the lens plane and source plane, respectively. For a finite source size (and/or outside the geometric optics assumptions), they effectively trace very high magnification regions \citep[e.g.,][]{1992grle.book.....S, 1996astro.ph..6001N}.
Indeed, the magnification factor for a given source depends on its position and size. Generally, the closer it is to the caustic in the source plane, and similarly, the smaller it is, the higher is the magnification factor ($\mu \propto D^{-1/2}$ or $\mu \propto R^{-1/2}$, with $D$ and $R$ being the source distance from the caustic and its size, respectively; \citealt{2017ApJ...850...49V}). For example, typical lensed galaxies are magnified by factors of a few to a few dozen, even if they lie on the caustic and form giant arcs in the image plane, whereas -- thanks to their small sizes -- stellar sources lying near the caustic can be extremely magnified by factors of $\sim 10^5$--$10^7$ \citep[e.g.,][]{1991ApJ...379...94M}.

\begin{figure*}
    \centering
    \includegraphics[width=18cm,height=13.5cm]{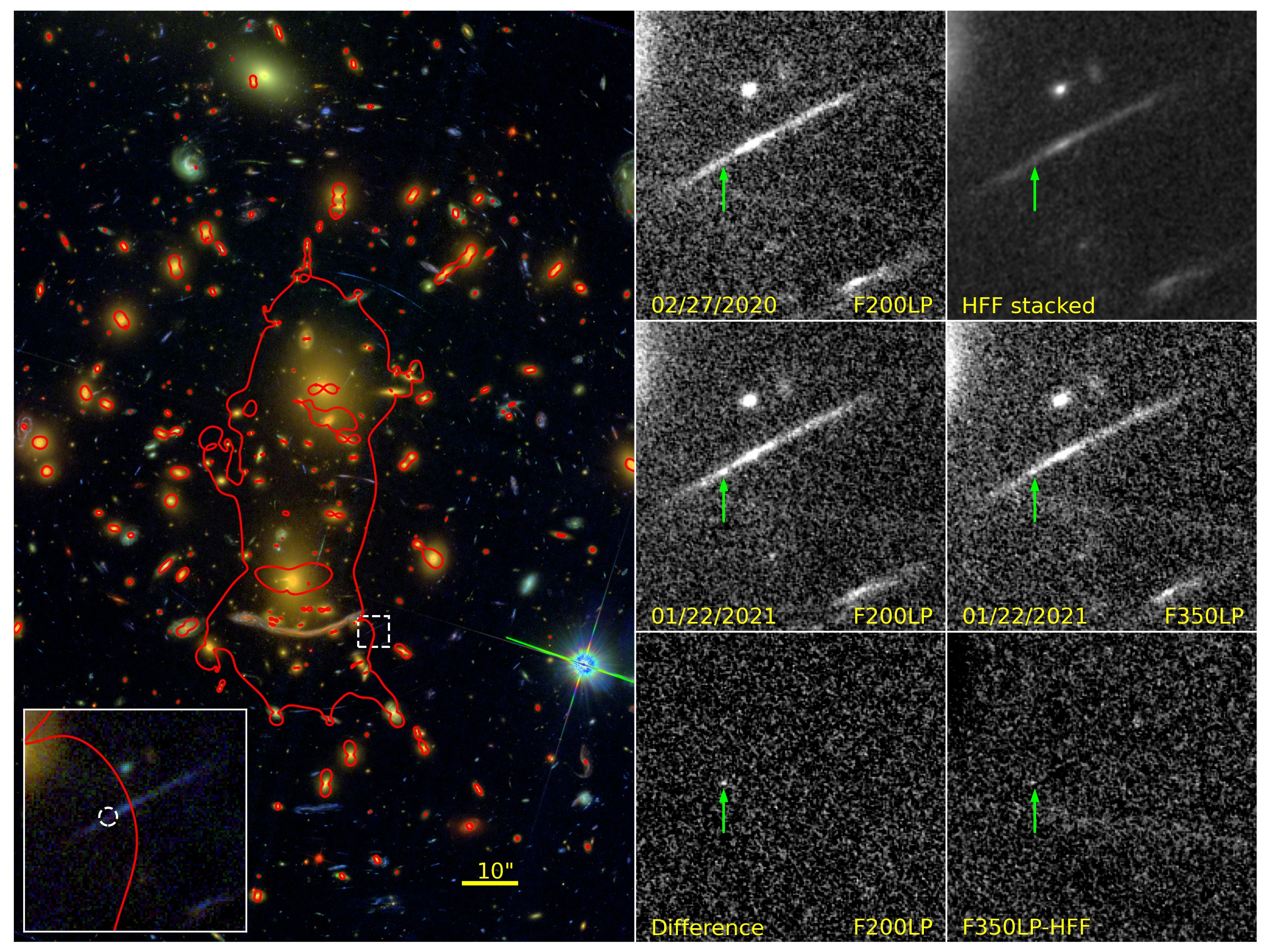}
    \caption{Left panel shows the $RGB$ image of the galaxy cluster Abell 370 created using ACS/WFC F435W, F606W, and F814W filters. The red line represents the critical curve corresponding to the \texttt{Glafic} model for source at $z_{\rm s}=1.2567$. The dashed white box represents the underlying arc at $z_{\rm s}=1.2567$ in which the transient was detected in the F200LP and F350LP filters. The zoomed-in version of the same white box is shown in the inset panel on the lower-left side. The middle column panels represent the F200LP filter observations in Feb. 2020 and Jan. 2021 and the difference image from top to bottom, respectively. The rightmost column represents a stacked \emph{HFF} image (made using F435W, F606W, and F814W filters), F350LP filter observation in Jan. 2021, and the corresponding difference image. Since F350LP filter observations were not available for Feb. 2020, the subtraction was done using the stacked \emph{HFF} image which effectively spans a similar wavelength range. In both the middle and right columns, the transient position is shown by the green arrows.}
    \label{fig:combPlot}
\end{figure*}

The presence of microlenses (e.g., stars, stellar remnants, and possible compact dark  matter objects) in the cluster lens leads to the formation of microcritical curves on small scales. The area covered by a microcritical curve depends, in addition to the microlens mass, on the macromagnification (i.e., magnification due to the cluster lens) at the microlens position \citep[e.g.,][]{2018ApJ...857...25D, 2018PhRvD..97b3518O}. Near the macrocritical curves as the macromagnification is very high, the microcritical curves merge with each other to form a corrugated network \citep[e.g.,][]{2017ApJ...850...49V, 2019A&A...625A..84D}. The width of this corrugated network of microcritical curves depends on the surface mass density of microlenses. In the source plane, this corrugated network results in a network of microcaustics. A sufficiently magnified star in the source plane that sweeps across this network of caustics, owing to the transverse motion between the lens, source, and observer, would show fluctuations in brightness, and often prominent peaks in its light curve. The observed width, height and frequency of these peaks depend on the source size, the transverse relative velocity between lens and source, the microlens mass functions, and microlens surface density.

In the past few years, several such lensed stars have been detected. \citet{2018NatAs...2..334K} reported the first-ever detection of a highly magnified star (a blue supergiant) in a spiral galaxy at $z=1.49$ in the field of the galaxy cluster MACS1149, marking the beginning of an era of observing stellar sources at high redshifts. The source star crossed a microcaustic, resulting in a peak in its light curve corresponding to a magnification of~$\gtrsim 2000$. In following years, several other transient sources were detected \citep{2018NatAs...2..324R, 2019ApJ...881....8C, 2019ApJ...880...58K, 2022arXiv220711658C, 2022A&A...665A.134D, 2022arXiv221006514D, 2023ApJ...944L...6M}, all the way up to a source redshift $z \sim6.2$ \citep{2022Natur.603..815W, 2022ApJ...940L...1W}\footnote{Strictly speaking, Earendel \citep{2022Natur.603..815W}, Godzilla \citep{2022A&A...665A.134D}, and Quyllur \citep{2022arXiv221006514D} were discovered based on the presence of a single image at the expected position of the critical curve, unlike previous examples or the current star candidate which were detected as transient sources.}. Many more such events are expected in the upcoming years, especially with the greater sensitivity of the \emph{James Webb Space Telescope (JWST)}. As the frequency of microlensing peaks (and their amplitudes) are sensitive to the microlens surface density and mass function, observing such caustic-crossing transients can allow us to put strong constraints on the fraction of compact dark matter in the cluster field \citep[e.g.,][]{2017ApJ...850...49V, 2018ApJ...857...25D, 2018PhRvD..97b3518O}. In addition, observed peak duration or alternatively magnification constraints also provide insight on the size of the source, adding valuable information about its nature.

In this paper, we report another detection of a transient in a strongly lensed arc at $z=1.2567$ in the Abell 370 galaxy cluster field. The transient is detected in WFC3/UVIS F200LP and F350LP images taken with the \emph{Hubble Space Telescope (HST)} in January 2021, through the \emph{Flashlights} program~\citep[PI Patrick L. Kelly;][]{2022arXiv221102670K}. Unlike other transients mentioned above, which are usually about $\sim0\farcs2 -\sim0\farcs3$ from the critical curves, the distance of the current transient from the inferred critical curve (passing through the lensed arc) is relatively large, $\gtrsim 0\farcs6$, resulting in a significantly lower background macromagnification. While here we do not have sufficient spectral coverage to perform a spectral energy distribution (SED) fit to independently classify the source and derive its properties, we show that the constraints from lensing together with the colour information are sufficient to rule out most other transient candidates, identify the source as -- most likely -- a star, and thus gather statistics about such events. In particular, if indeed this is a lensed star as suggested, it implies that such events will be seen in a substantial area around the critical curves, meaning many more of these should be detected soon in dedicated cluster surveys. 

This paper is organised as follows. In Section~\ref{sec:data} we describe the imaging data used to detect the transient. Details regarding the lens mass models of the Abell 370 galaxy cluster are presented in Section~\ref{ssec:models}. The nature of the transient is discussed in Section~\ref{ssec:nature}. Our conclusions are presented in Section~\ref{sec:conclusions}. Throughout this work, we use AB magnitudes \citep[][]{okegunn83}. The cosmological parameters used in this work to calculate the various quantities are H$_0 = 70 \: {\rm km \: s^{-1} Mpc^{-1}}$, $\Omega_\Lambda =0.7$, and $\Omega_m = 0.3$.

\begin{figure}
    \centering
    \includegraphics[width=8.5cm,height=8.5cm]{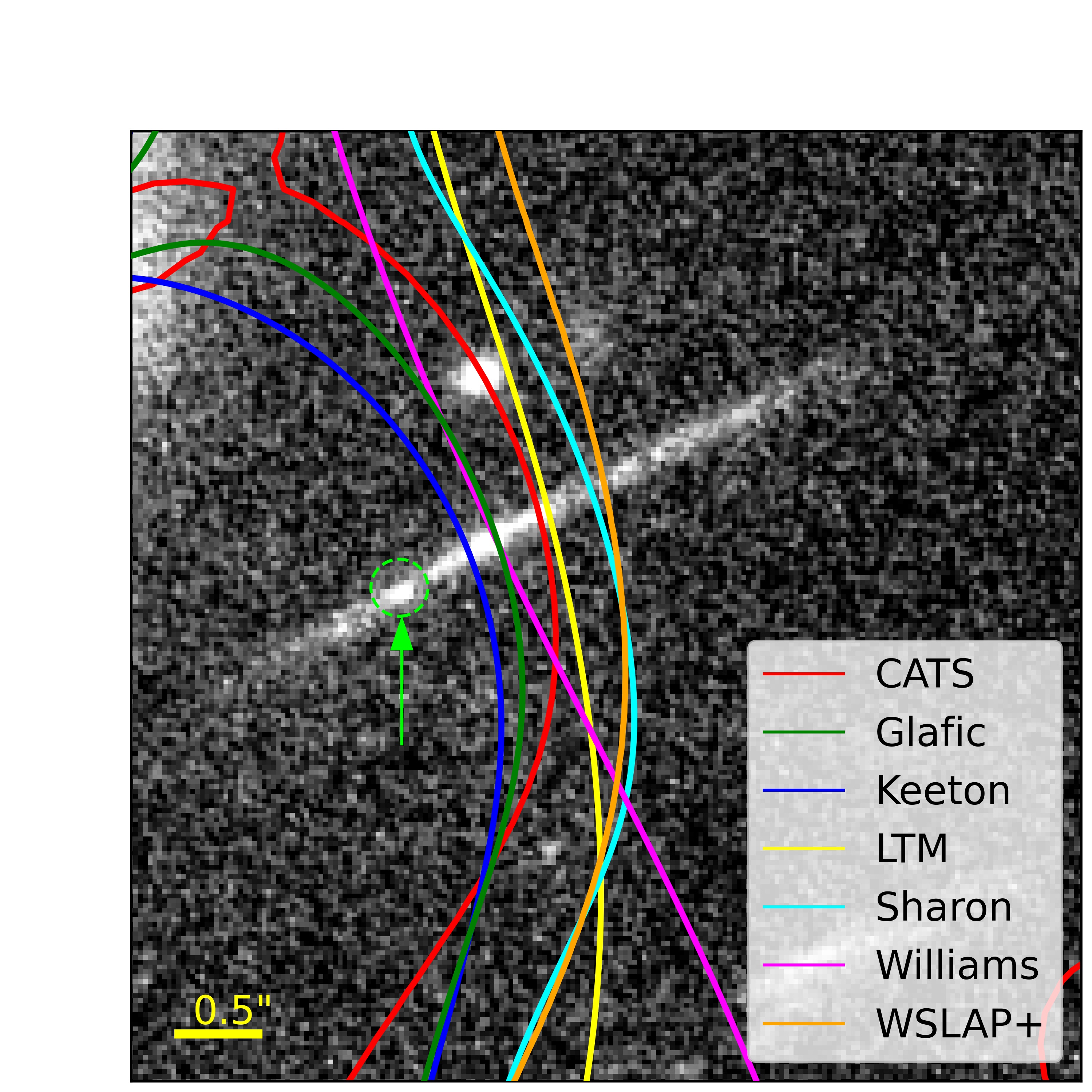}
    \includegraphics[width=8.5cm,height=8.5cm]{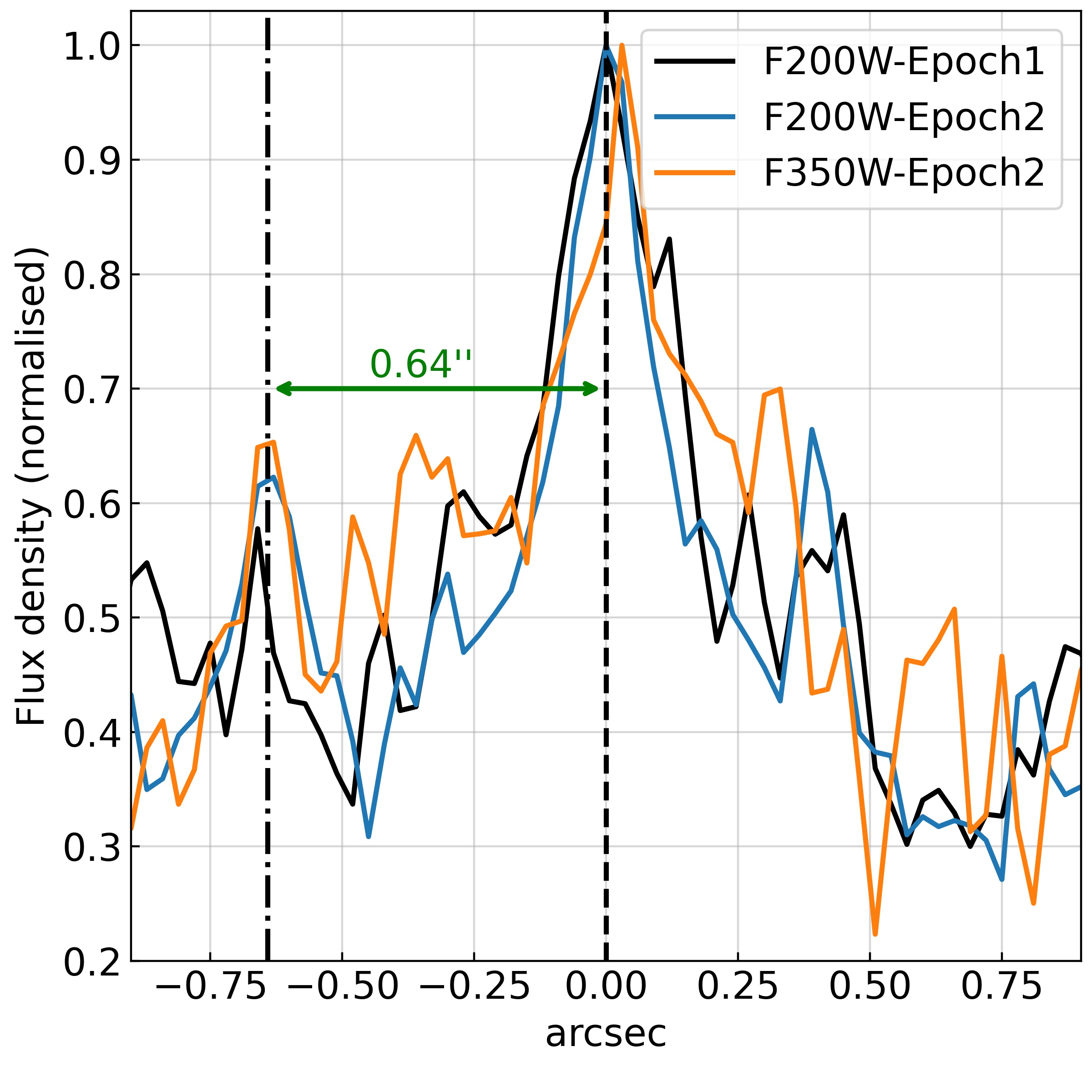}
    \caption{{\it Top panel:} critical curves passing through the relevant lensed arc at $z_{\rm s}=1.2567$ for different HFF-v4 mass models. The red, green, blue, yellow, cyan, magenta, and orange lines represent the critical curves for \texttt{CATS}, \texttt{Glafic}, \texttt{Keeton}, \texttt{LTM}, \texttt{Sharon}, \texttt{Williams}, and \texttt{WSLAP+} group mass models, respectively. The transient position in the F200LP filter during the second epoch is circled and pointed by a green arrow. {\it Bottom panel:} one-dimensional (1D) light profile along the arc. The black, blue, and orange curves represent the F200W-Epoch1, F200W-Epoch2, and F350W-Epoch2 1D normalised light profiles. The vertical black dashed line marks the expected position of the critical curve. The black dashed-dotted vertical line represents the transient position along the arc. The distance between the expected position of the critical curve and the transient position is $0\farcs64$.}
    \label{fig:criticalCurve}
\end{figure}

\begin{figure*}
    \centering
    \includegraphics[scale=0.4]{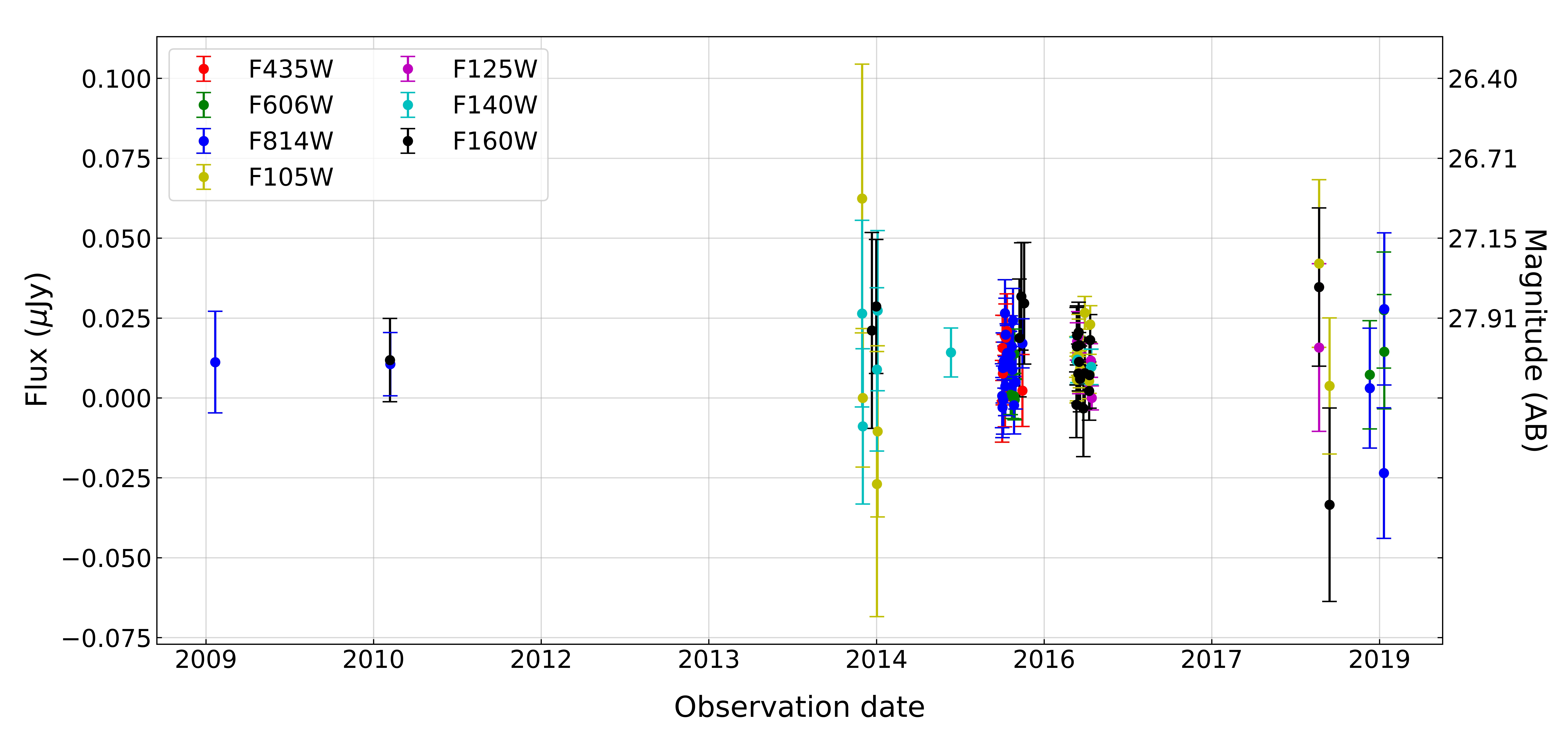}
    \caption{Photometry at the transient position in the archival \emph{HST} images of Abell 370 with an aperture size of $0\farcs2$. The error bars are for flux values only. We stress that this plot is based on earlier \emph{HFF} observations and can only be used to infer the (lack of) variability at the \emph{HFF} detection level. We cannot use it to rule out a fainter variable source at the \emph{Flashlights} detection level.}
    \label{fig:ObslightCurve}
\end{figure*}

\section{Data}
\label{sec:data}
 
Imaging of Abell 370 has been acquired under the \emph{Flashlights} program, which targets all of the \emph{Hubble Frontier Fields} \citep[\emph{HFF;}][]{2017ApJ...837...97L} clusters with \emph{HST} at separate epochs. The aim of \emph{Flashlights} is to build up a statistical sample of microlensing events, observed in strongly lensed merging images or arcs in galaxy clusters, to better understand the intracluster medium properties and high-redshift ($z>1$) Universe. \emph{Flashlights} is a two-epoch program (using the ultrawide F200LP and F350LP long-pass WFC3/UVIS filters), with each visit reaching a 5$\sigma$ limiting mangitude of~$\sim 31$~AB,~$\sim 2$~mag deeper than the existing \emph{HFF} images.

The left panel in Figure~\ref{fig:combPlot} shows an RGB image of Abell 370 (made using \emph{HST} ACS/WFC F435W, F606W, F814W filters), highlighting and zooming-in on the strongly lensed arc at $z=1.2567$~\citep{2017MNRAS.469.3946L} in which the transient was detected. The transient J2000 position ($\alpha$, $\delta$) = ($2^{\rm hr}39^{\rm m}52.1308^{\rm s}$, $-1^\circ 35' 05\farcs755$) in the world coordinate system (WCS) is marked by the white-dashed circle in the inset image. The middle column represents the \emph{HST} visits on 2020 Feb. 27 and 2021 Jan. 22 (UT dates are used throughout this paper), and their difference image in the WFC3-F200LP filter, from top-to-bottom, respectively. Unfortunately, owing to some technical issues, we only have a WFC3-F350LP filter image corresponding to the second-epoch observations. Hence, we made a stacked image using earlier \emph{HFF} observations in F435W, F606W, and F814W (combined together, these filters cover a wavelength range similar to that of the F350LP filter), and subtracted it from the second-epoch WFC3-F350LP filter image. The right column represents the \emph{HFF} stacked image, the WFC3-F350LP image taken on 2021 Jan. 22, and their difference image (i.e., the \emph{HFF} stacked image subtracted from the F350LP image) from top to bottom. The transient position is shown by the green arrows in both the middle and right columns. The transient is visible in both the F200LP and F350LP filter difference images.

We re-sampled the two-epoch \emph{HST} imaging to a scale of $0\farcs03~\mathrm{pixel}^{-1}$ using {\tt AstroDrizzle} after aligning the images using {\tt TweakReg} \citep{drizzlepac}. We then used {\tt PythonPhot} \citep{pythonphot} to measure the photometry of the transient from the difference between the images from the two epochs. The {\tt PythonPhot} package provides tools to fit a point-source flux density for the transient in the difference imaging based on the {\tt DAOPHOT} algorithm \citep{daophot}. We obtain an apparent magnitude for the transient of $29.51\pm0.14$~AB from the F200LP difference image and~$\sim 30.53\pm0.76$~AB from the~F350LP-HFF difference image~(see Section~\ref{ssec:cce} for more details). 
To measure the intracluster light~(ICL) surface mass density of stars in the vicinity of the arc near the lensed star, we measure the light in two rectangular apertures on the two sides of the lensed arc. The measurement is based on the archival \textit{HST} ACS/WFC F435W, F606W, F814W and WFC3/IR F105W, F125W, F140W, and F160W imaging acquired from the HFF program~\citep{2017ApJ...837...97L}. Based on the median pixel fluxes within the two apertures, we fit the spectral energy distribution (SED) using the \texttt{FAST++} package\footnote{\url{https://github.com/cschreib/fastpp}}, a \texttt{C++} version of the Fitting and Assessment of Synthetic Templates~(\texttt{FAST}) code~\citep{2009ApJ...700..221K}.
Based on this, we obtain ICL surface mass density near the position of the transient to be $4.37^{+0.76}_{-1.62}~{\rm M}_\odot$~pc$^{-2}$. The total stellar mass and star formation of the arc (without magnification correction) are $1.28 \times 10^8~{\rm M}_\odot$ and $0.24~{\rm M}_\odot~{\rm yr}^{-1}$, respectively \citep{2018ApJS..235...14S}\footnote{The values of stellar mass and star formation in the arc are taken from the HFFDeepSpace website:~\url{http://cosmos.phy.tufts.edu/~danilo/HFF/HFFexplorer.html}}.

\section{Lens Models of Abell 370}
\label{ssec:models}

Under the \emph{Hubble Frontier Fields} initiative\footnote{\url{https://archive.stsci.edu/prepds/frontier/}}~\citep[e.g.,][]{2017ApJ...837...97L}, various groups modeled Abell 370 using different reconstruction methods \citep[e.g.,][]{2017MNRAS.472.3177M}. The critical curves near the underlying arc ($z=1.2567$; \citealt{2017MNRAS.469.3946L}) for seven different lens models are shown in Figure~\ref{fig:criticalCurve}. The red, green, blue, yellow, cyan, magenta, and orange colours represent the critical curves corresponding to \texttt{CATS}~\citep[e.g.,][]{2014MNRAS.444..268R}, \texttt{Glafic}~\citep[e.g.,][]{2010PASJ...62.1017O, 2018ApJ...855....4K}, \texttt{Keeton}~\citep[e.g.,][]{2014MNRAS.443.3631M}, \texttt{LTM}~\citep[e.g.,][]{2009MNRAS.396.1985Z} \texttt{Sharon}~\citep[e.g.,][]{2014ApJ...797...48J}, \texttt{Williams} \citep[e.g.,][]{2019MNRAS.488.3251S}, and \texttt{WSLAP+} \citep[e.g.,][]{2018MNRAS.473.4279D} best-fit mass models, respectively. The light-green arrow and dashed circle highlight the transient position. The background is the F200LP observation during the second epoch.

The arc is a typical example of a source lying on a fold caustic in the source plane, giving rise to two merging images (with mirror symmetry) along with a third image on the other side of the cluster. From the top panel in Figure~\ref{fig:criticalCurve}, we can see that the position of the critical curve near this arc is not well constrained, as the scatter in the critical curves corresponding to different mass models is quite large ($\gtrsim 0\farcs5$). However, we can still deduce the centre of the arc to be at the centre of the bright spot following the mirror-symmetry argument.\footnote{Otherwise, we would expect a counterimage of the bright spot on the other side of the macrocritical curve, which is not observed.}. Focusing on the apparent centre of the arc and the critical curves corresponding to the \texttt{Glafic}, \texttt{Keeton}, and \texttt{Williams} mass models which pass very near to this apparent centre, we can bring down the uncertainty in the critical curve position to $\sim 0\farcs2$. Since these three models agree best with the apparent symmetry of the arc, we only consider them. The transient appears on the inner side (i.e., negative parity side) of the critical curve and the distance between the transient position and the critical curve for \texttt{Glafic}, \texttt{Keeton}, and \texttt{Williams} models is $\sim 0\farcs67$, $\sim 0\farcs48$, and $\sim 0\farcs67$, respectively. In comparison with the distance of the highly magnified stars detected in earlier works \citep[$0\farcs1$--$0\farcs3$; e.g.,][]{2018NatAs...2..334K, 2019ApJ...881....8C, 2022Natur.603..815W}, the distance of the present transient from the critical curve is significantly large. The macromagnification ($|\mu|$) values corresponding to these three mass models at the transient position are $\sim45$, $\sim65$, and $\sim35$, respectively\footnote{The distance from the critical curve (magnification) values of the present transient according to \texttt{CATS}, \texttt{LTM}, \texttt{Sharon}, and \texttt{WSLAP+} lens models are $0\farcs99~(42)$, $1\farcs05~(39)$, $1\farcs35~(20)$, and~$1\farcs44~(32)$, respectively.}. The observed time delay between the transient and its expected counterpart on the arc (on the other side of the critical curve), which is not observed, is around $\sim0.8$ days, $\sim0.3$ days, and $\sim1$ days (and around~30~years with respect to the global minimum image on the other side of the cluster) according to the \texttt{Glafic}, \texttt{Keeton}, and \texttt{Williams} mass models. 

To further decrease the uncertainty in the critical curve position, we plot the 1D light profile along the arc in the bottom panel in the F200W observation in Feb. 2020 (black curve), the F200W observation in Jan. 2021 (blue curve), and the F350W observation in Jan. 2021 (orange curve). The peak in the light profile (black dashed curve) denotes the expected position of the critical curve along the arc. The transient position in the light profile is shown by the black dashed-dotted curve. The distance to the transient from the critical curve is $0\farcs64$, which agrees very well with the \texttt{Glafic} and \texttt{Williams} models. Assuming that the critical curve corresponding to \texttt{Glafic} passes through the centre of the arc, to improve the macromagnification estimate we determine the macromagnification at a distance of $0\farcs67$ from the critical curve for other lens models resulting in a value $|\mu| \sim [40, 50]$, which is very close to the \texttt{Glafic} mass model value. Hence, in the following sections, we use parameter values corresponding to the \texttt{Glafic} mass model for various estimations, including the macromodel magnification.

\section{Nature of the Transient}
\label{ssec:nature}

From Figure~\ref{fig:combPlot}, we can see that the transient was visible in both F200LP and F350LP in the second epoch, but not in the first epoch (where for F350LP as mentioned we used a post-processed combined HFF image instead of the first epoch). In addition, as the source lies near (and inside) a fold caustic, it will lead to formation of two images (one inside and one outside of the critical curve) with an observed time delay of $\lesssim0.8$ days following the \texttt{Glafic} model, with the detected image being the trailing image (arriving last). As we do not detect any image that could be a counter image on the other side of the critical curve in the difference images, we can eliminate any transient lasting more that $\sim0.8$~days in the observer frame (and $\sim0.35$~days in the source rest frame) as the possible source candidate. The above information does not allow us to determine the exact nature of the transient, but we can still substantially narrow down the list of viable source candidates as discussed in the following subsections.

\begin{table}
    \caption{Likelihood ratio test in different filters to estimates the variability in the light curve. Here we have assumed that the mean flux density of the light curve in a given filter remains constant over the years as the \emph{null hypothesis}. The alternative hypothesis is that the mean changes with time. The first column shows the name of the \emph{HST} filter and second column gives the corresponding \emph{degrees of freedom} (i.e., the number of time segments into which the light curves is divided). The third column represents the $\chi^2$ value from the likelihood ratio test, and the fourth column is the corresponding p-value.}
    \label{tab:LRTest}
    \begin{tabular}{ccccc}
        \hline
        Filter Name & Degrees of Freedom & $\chi^2$ & p-value \\
        \hline
        F435W & 2 & 0.85 & 0.65336 \\
        F606W &	2 & 1.61 & 0.44669 \\ 
        F814W & 3 & 0.73 & 0.86501 \\
        F105W &	3 & 0.85 & 0.83765 \\
        F125W & 2 & 0.08 & 0.95985 \\
        F140W & 4 & 0.46 & 0.97734 \\
        F160W &	5 & 3.00 & 0.69995 \\
        \hline
    \end{tabular}
\end{table}

\subsection{Variable Source}
\label{ssec:lbv}

To quantify the level of variability at the transient position, we perform a likelihood ratio test in all bands shown in Figure~\ref{fig:ObslightCurve} (for more details, see Zhou et al., in prep.). If the source is not variable, then the corresponding light curve in different filters will only show random fluctuations around a mean value, within the uncertainties. Hence, we assume that each observation is a random variable following a normal distribution around a mean value as our \emph{null hypothesis}. Alternatively, if the source is indeed variable, then different data points at different times in the light curve will follow a normal distribution but around different mean values. We compare these two hypotheses using the likelihood ratio test and estimate the $\chi^2$ value based on that. We show the results in Table~\ref{tab:LRTest}. The first column represents the different filters, and the second column shows the number of degrees of freedom (DoF). In our case, the DoF essentially represent the number of segments into which the light curve was divided for the second hypothesis assuming that the corresponding mean value changes over time. The third and fourth columns represent the corresponding $\chi^2$ value and the p-value (i.e., significance level). A p-value $<0.05$ implies that we can rule out the null hypothesis with a confidence level of 95\%. From Table~\ref{tab:LRTest}, we can see that for each filter the p-value is significantly larger than 0.05, meaning that the null hypothesis is preferable and cannot be ruled out at the 95\% confidence level. Hence, for earlier \emph{HFF} observations, we can exclude the possibility of any previous variability at the transient position. 


We note that the above analysis only implies that we do not find any evidence in favour of variability at the \emph{HFF} survey detection level (depth of an individual observation $\sim 26.3$--27~AB; depth of the coadded images in the different filters $\sim 28$--29~AB; \citealt{2017ApJ...837...97L}). We cannot dismiss a variable source at the \emph{Flashlights} survey detection limit, which is $\sim2$~mag deeper than the coadded depth of the \emph{HFF}. However, as mentioned above, the rest-frame timescale for the transient to fade below the \emph{Flashlights} detection level is $\sim0.35$~days i.e., $\sim0.8/(1+z_s)$~days (otherwise we would have also detected the counterimage), putting a strong upper bound on the outburst duration. As discussed by \citet{2018NatAs...2..324R}, a typical luminous blue variable (LBV) source may take several years to rise by $\sim 2$~mag. However, rapid outbursts in LBV sources were also observed, with rest-frame timescales of the order of days for the source to become fainter by a factor of $\sim 100$ (see Figure~5 of \citealt{2018NatAs...2..324R}). One example of such a short outburst is discussed by \citet{2010MNRAS.408..181P}, where the peak luminosity was followed by a steep decline ($\sim3$~mag decrease in $\sim2$ days). Hence, a rapid LBV outburst which happened in the source plane might in principle explain the observed transient, but the scarcity of such known examples makes the LBV origin very unlikely. In addition, as we have observations at two epochs separated by a year, any LBV in the lens plane with a an observed duration of less than a year is thus also a valid candidate, in principle.

\subsection{Kilonova or Nova}
\label{ssec:nova}

As the transient was detected at an apparent AB magnitude of $29.51{\pm}0.14$ in F200LP, assuming that the transient lies in the source plane the corresponding (magnification corrected) luminosity in F200LP is expected to be $\sim1.1\times10^{40}$~erg~s$^{-1}$ and the absolute AB magnitude is $-11.07$ (without K-correction). On the other hand, if we assume that the transient lies in the lens plane, then its F200LP luminosity and absolute AB magnitude are $\sim2.7\times10^{40}$~erg~s$^{-1}$ and $-12.01$ (without K-correction), respectively.

A kilonova is the result of a merger of two neutron stars with a peak luminosity of $\sim10^{40}$--$10^{41}$~erg~s$^{-1}$, thereby satisfying the above luminosity requirement. For a kilonova, the rest-frame timescale is about one week for its brightness to decline by 2~mag \citep[e.g.,][]{2016AdAst2016E...8T, 2020ApJ...889..171K}. As the observed transient is the trailing image with an observed time delay of $\sim0.8$~days, we can rule out a kilonova from the possible list of candidates in the source galaxy; otherwise, we should have observed the counterpart on the minima side of the critical curve. However, we cannot rule out a kilonova in the lens plane, as the two observations are separated by $\sim 1$~yr, although the scarcity of such events again makes chance alignment with the arc very unlikely.

Another possible candidate is a nova, a sudden outburst resulting from mass transfer from a star to a white dwarf in a binary system \citep[e.g.,][]{2020JApA...41...43A, 2021ARA&A..59..391C}. At peak, the luminosity is  $\sim10^{38}$--$10^{40}$~erg~s$^{-1}$, with the rest-frame characteristic timescale to decline by 2~mag being 10--100~days. The typical time between two consecutive outbursts in such novae can be 10--100~yr. Again, from the time-delay constraint, we can rule out a nova lasting longer than the observed duration of~$0.8$ days. In addition, from the \emph{maximum magnitude vs. rate of decline} relation (MMRD; \citealt{1995ApJ...452..704D}) for classical novae (see also Figure~5 of \citealt{2018NatAs...2..324R}), the maximum allowed absolute AB magnitude for a nova with a duration of $\sim0.35$~days is $\sim -11$. The estimated absolute AB magnitude for the transient in F200LP is $-11.07$ (~$\sim-12$ if k-corrected) or $-12.01$ if the transient lies in the source or lens plane, respectively, implying that if the transient were a nova, the corresponding peak luminosity nearly equals (or is greater than) the peak luminosity allowed from the MMRD relation. Hence, considering both the allowed duration of the transient and its luminosity, it is very unlikely to be a nova in the source plane. For a nova in the lens plane, an absolute AB magnitude value of $-12$ makes it a very luminous nova (and a very short one) according to the MMRD relation, again making it a very unlikely. 

In addition to the above arguments, note that the colour of the transient estimated using the F200W and F350W filters (i.e., $M_{\rm F200W}$--$M_{\rm F350W}$; see Table~\ref{tab:StellarExp}) discussed in Section~\ref{ssec:cce} is very blue for kilo(nova) sources \citep[e.g.,][]{2016A&A...592A.134T, 2019ApJ...878...28S, 2022MNRAS.515..631G}. Hence, considering the time-delay constraint and the colour of the transient, we can rule out the kilo(nova) from the list of transient candidates.

\begin{table*}
    \caption{Apparent magnitudes of different types of stars in F200LP and F350LP. Column~1 represents the spectral type of the star whereas Columns~2 and 3 show the corresponding temperature and radius, respectively. Columns~4 and 5 represent the apparent magnitude in F200LP and F350LP assuming that the star is at $z_s = 1.2567$. Column~6 is the magnitude difference between F200LP and F350LP. Column~7 represents the magnification needed for the star to be visible at an AB magnitude of 29.5 in F200LP. The final Column~8 represents one example star of the spectral class. For these stars, the SEDs were take from \citet{2003IAUS..210P.A20C}.}
    \label{tab:StellarExp}
    \begin{tabular}{cccccccc}
        \hline
        Spectral type & Temperature & Radius & $m_{200}$ & $m_{350}$ & $\Delta m$ & $\mu_{200}$ & Comments \\
             (1)      &    (2)      &  (3)   &    (4)    &   (5)     &    (6)     &    (7)      &    (8)   \\
        \hline
        A0V & 9500~K   &    3~${\rm R}_\odot$ & 46.10 & 45.61 & 0.49  & $4.3 \times 10^6 $ & Vega; main sequence \\
        B8V  & 12,000~K &   79~${\rm R}_\odot$ & 37.61 & 37.24 & 0.37  & $1.7 \times 10^3$ & Rigel; blue supergiant \\
        B0V  & 30,000~K &  6.5~${\rm R}_\odot$ & 38.92 & 39.10 & -0.18 & $5.9 \times 10^3$ & $\tau$~Scorpii; main sequence \\
        O3V  & 45,000~K &   12~${\rm R}_\odot$ & 36.50 & 36.82 & -0.32 & $6.3 \times 10^2$ & Main sequence \\
        \hline
    \end{tabular}
\end{table*}

\begin{figure}
    \centering
    \includegraphics[scale=0.57]{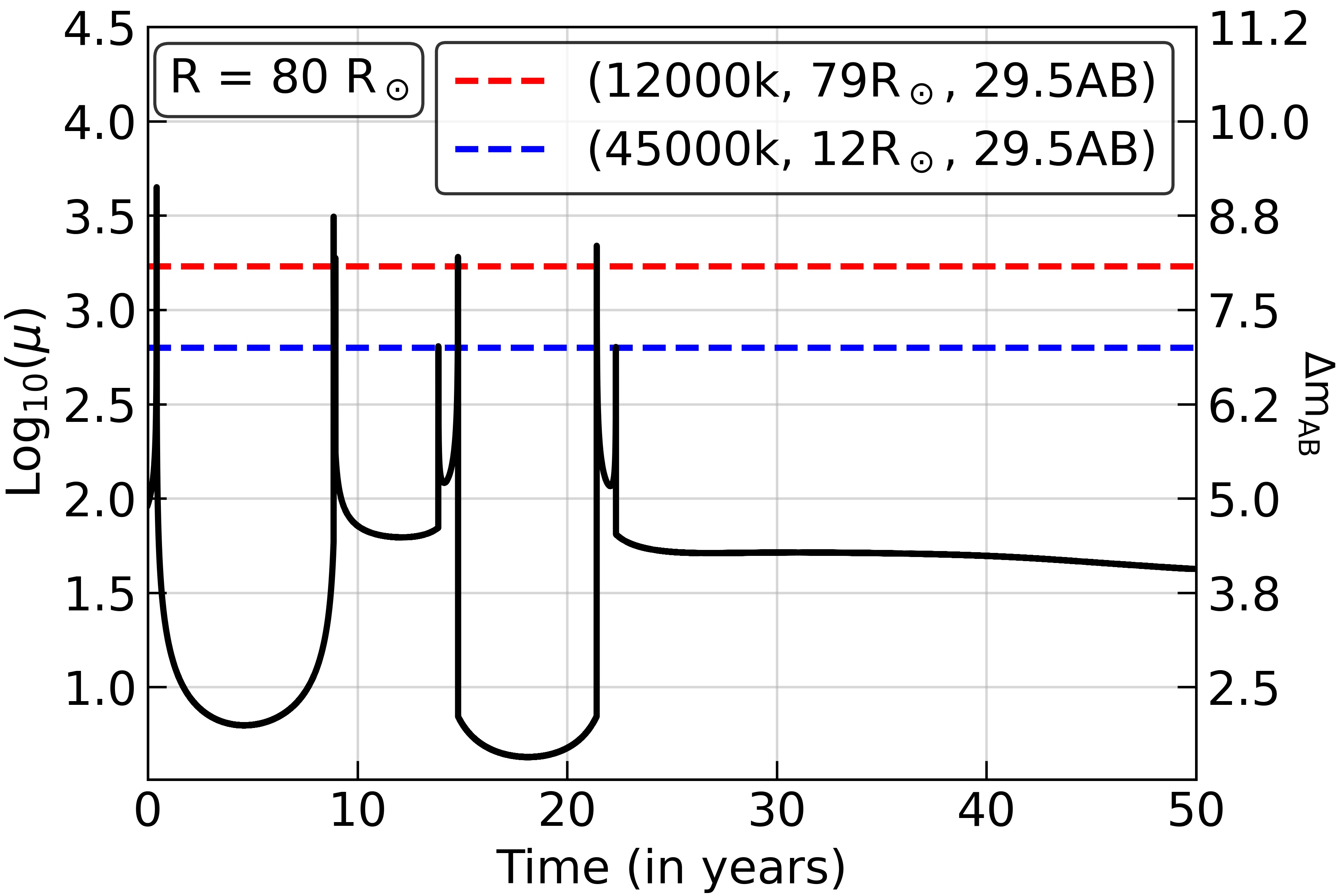}
    \caption{An example of a microlensing light curve at the transient position. The convergence and shear values were taken from the \texttt{Galfic} model. The microlens density is assumed to be $5~{\rm M}_\odot$~pc$^{-2}$ with microlens masses in the mass range [0.08, 1.5]~${\rm M}_\odot$, a Salpeter mass function, and a source radius of $80~{\rm R}_\odot$. The abscissa represents the observation time and the left-hand-side ordinate represents the logarithmic value of the magnification factor, ${\rm Log}_{10}(|\mu|)$. The right-hand-side ordinate shows the magnitude difference with respect to the unlensed case. The blue and red lines represent the magnification needed for O- and B-type stars with a respective temperature of 12,000~K and 45,000~K, and with a respective radius 79~${\rm R_\odot}$ and 12~${\rm R_\odot}$, to reach an apparent AB magnitude of 29.5.}
    \label{fig:SimlightCurve} 
\end{figure}

\begin{figure}
    \centering
    \includegraphics[width=8.6cm,height=6.2cm]{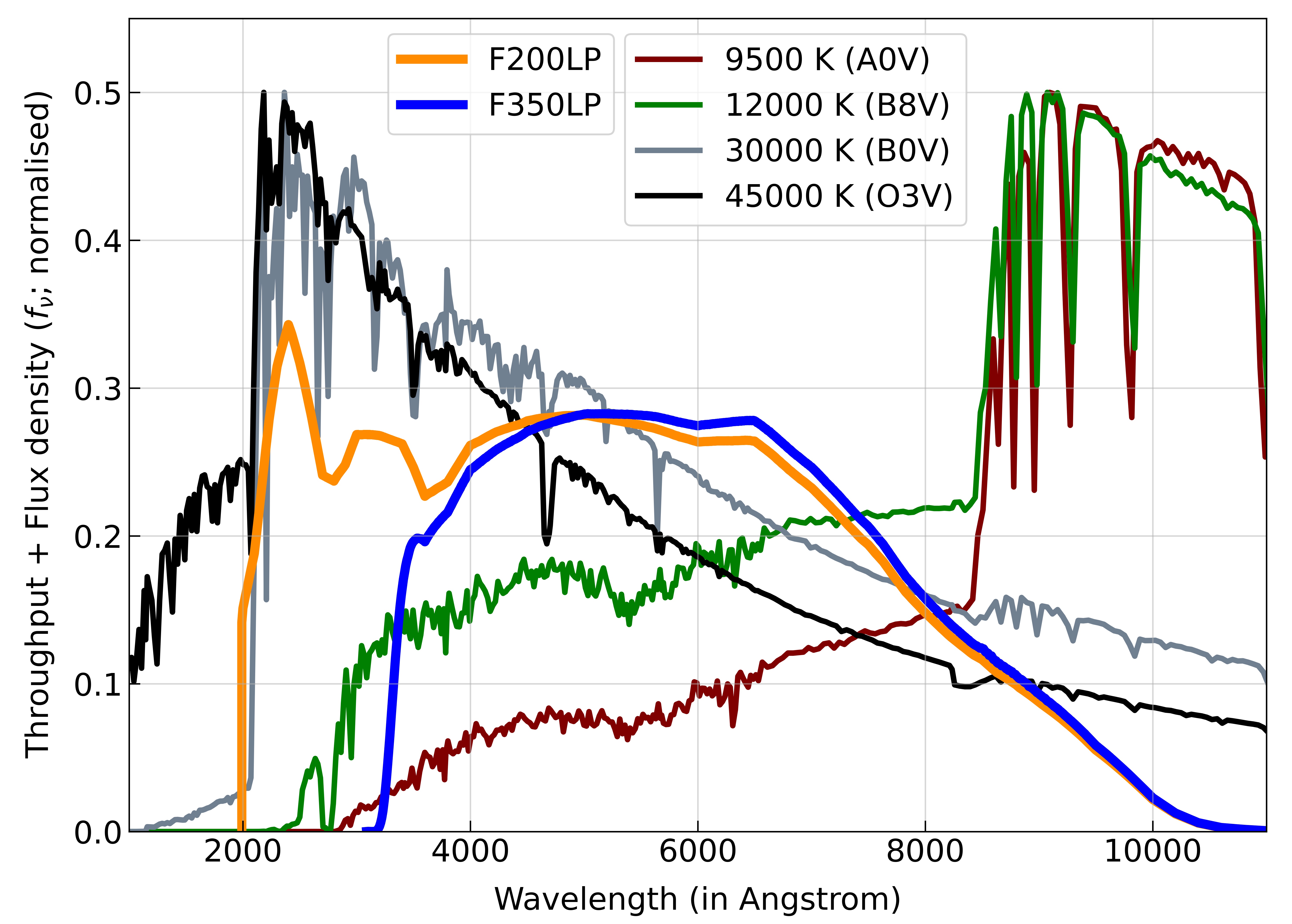}
    \caption{F200LP and F350LP filter throughput curves along with four different stellar SEDs. The blue and orange curves represent the throughput corresponding to F200LP and F350LP, respectively. The maroon, green, cyan, and black colour curves show the observed SEDs of stars at the redshift of the arc ($z=1.2567$) with temperatures values 9500~K, 12,000~K, 30,000~K, and 45,000~K, respectively.}
    \label{fig:FilterSED}
\end{figure}

\subsection{Highly Magnified Star}
\label{ssec:cce}

The most interesting candidate for the transient is the microlensing of an O/B-type star. An O-type main-sequence star with radius $\sim 12~{\rm R}_\odot$ and temperature $T \sim 45,000$~K needs a magnification factor $\sim650$ to be visible at 29.5~AB~mag at the transient position. On the other hand, a blue supergiant (B-type) star like Rigel with radius $\sim 79~{\rm R}_\odot$ and temperature $T \sim 12,000$~K needs a magnification of $\sim1700$ to be visible at 29.5~AB\footnote{Here, SEDs and corresponding temperatures for O- and B-type stars are taken from \citet{2003IAUS..210P.A20C}. Also, see Tables~1 and 2 of \citet{2022MNRAS.514.2545M} for an estimate of the magnification needed to see different types of stars in various \emph{HST} and \emph{JWST} filters.}. An example of a light curve for a star of $80~{\rm R}_\odot$ with a typical microlens density of $5~{\rm M}_\odot$~pc$^{-2}$ is shown in Figure~\ref{fig:SimlightCurve} using the convergence ($\kappa=0.702$) and shear ($\gamma=0.333$) value at the transient position from the \texttt{Glafic} mass model. We used a Salpeter mass function in the range [0.08, 1.5]~${\rm M}_\odot$ to generate the microlens population. The light curve is simulated using the \emph{adaptive boundary method} \citep[\texttt{ABM};][]{2022MNRAS.514.2545M}, which is designed to produce accurate high-resolution ray-tracing light curves of small sources in an efficient way. The abscissa represents the time in years whereas the left-side ordinate represents ${\rm Log}_{10}(|\mu|)$. The right-side ordinate shows the gain in apparent magnitude ($\Delta m_{\rm AB}$) compared to the unlensed case, $-2.5\:{\rm Log}_{10}(|\mu|)$. From Figure~\ref{fig:SimlightCurve}, we note that a stellar source with radius $80~{\rm R}_\odot$ can lead to peaks with $|\mu|>3000$ in the light curve, implying that O/B-type stars are valid source candidates for the present transient. As the transient appears on the saddle side of the critical curve, we also note the regions with significant de-magnification (near $\sim15$~yr and $\sim45$~yr in the light curve). Typically, peaks appearing before and after these demagnified regions are stronger in order to conserve the total flux over long time periods \citep{2018ApJ...857...25D}. As the peak magnification is inversely proportional to the square root of the source radius (i.e., $\mu \propto 1/\sqrt{R}$), we can easily estimate the peak magnification for a source with radius $R$ by multiplying the peak values by $\left(80~{{\rm R}_\odot}/R\right)^{1/2}$.

As the F200LP and F350LP filters cover different wavelength ranges, the apparent magnitude of the transient in these filters can offer useful information to constrain the SED of the transient. For F200LP, we subtract the first epoch image from the second epoch image, and measure a $29.51\pm0.14$~AB from the difference image. However, for the underlying transient, we do not have an F350LP observation in the first epoch. Hence, keeping in mind that we do not observe any variability at the transient position in \emph{HFF} observations, we make a stacked image using F435W, F606W, and F814W \emph{HFF} observations to compare the apparent magnitude of the transient in F200LP and F350LP. This effectively creates a very broad-band filter, roughly as wide as the F350LP band. However, as the wavelength sensitivity of the \emph{HFF} stacked image (i.e., its effective filter response) differs from that of the F350LP filter, subtracting the stacked image from the latter may leave a significant residuals depending on the colours of the object in question. To remove these residuals, we subtract the stacked image multiplied by a constant factor defined such that the root-mean square (rms) at the arc position is minimum outside the $0\farcs15$ radius circle situated on the transient. We find that this procedure works very well, owing to the relatively uniform colour across the arc, with the residual rms of the same order of that obtained in the F200LP subtraction. We measure~$\sim 30.53\pm0.76$~AB from the~F350LP-HFF difference image, so that the transient apparent $m_{\rm F200LP}-m_{\rm F350LP}$ colour is quite blue, in the range $\in [-0.3, -1.6]$ AB mag.

In Figure~\ref{fig:FilterSED}, we plot the SEDs of four different types of stars along with the F200LP and F350LP filter throughputs. The corresponding magnitudes in these two filters and their differences are shown in Table~\ref{tab:StellarExp}. We notice that a blue supergiant like Rigel has a positive (i.e., redder) colour, implying that it will be brighter in F350LP than in F200LP. On the other hand, $\tau$~Scorpii (a massive OB star) and O3~V main-sequence stars show negative colours, consistent with the above subtracted images. This behaviour can be understood from Figure~\ref{fig:FilterSED}, where the flux corresponding to a 12,000~K star declines significantly below 2500~\AA, leading to less flux in F200LP. On the other hand, B0~V and O3~V stars at the arc redshift peak around 2000~\AA, making them brighter in F200LP. Hence, a massive O-type star is more preferred than a blue supergiant as a candidate.

Figure~\ref{fig:SimlightCurve} shows that even the typical microlens population made of ordinary-mass stars in the range $[0.08, 1.5]~{\rm M_\odot}$ is capable of giving rise to peaks that are bright enough to make O/B-type stars observable at the transient position. However, the frequency of these peaks is low ($\sim1$~per 10~yr, per source star; this is roughly also the rate in which we expect to see caustic crossings at the transient position). The frequency of these can be increased by increasing the microlens surface mass density\footnote{Although beyond some threshold value, higher microlens density will cause the light curve to saturate, when at each point several caustics will intersect the star.}, but this will decrease the maximum peak magnification. The frequency of these peaks can also be increased in the presence of massive microlenses (i.e., intermediate-mass black holes) or subhalos (in the mass range [$10^6$--$10^8$]~M$_\odot$) near the transient position \citep[e.g.,][Williams et al., in prep.]{2018ApJ...857...25D}. Introducing microlenses near the subhalo critical curve will lead to the formation of microcritical curves with peak values higher than the no-subhalo case. A microlensing event far from the macrocritical curve ($\sim1''$) might thus give us a strong indication for subhalo lensing.

\section{Conclusions}
\label{sec:conclusions}
We report the discovery of a transient in a strongly lensed arc ($z=1.2567$) in the Abell 370 galaxy cluster, observed as part of our \emph{Flashlights} program. The transient was detected in the difference image made using the ultra-wide {\it HST} F200LP filter with an apparent AB magnitude of $29.51\pm0.14$, and is also observed at $\sim 30.53\pm0.76$ in the F350LP filter. Compared to previously observed caustic transients, it lies at a larger distance ($\gtrsim0\farcs6$) from the critical curve passing through the arc, with a modest background macromagnification $|\mu| \sim 45$, which we find is sufficient for bright supergiants to be observed during micro-caustic transits. 

We have performed a $\chi^2$ analysis using earlier \emph{HFF} observations to detect any variability at the transient position. We do not find any evidence of variability in the \emph{HFF} observations at the transient position, implying that there is no variable LBV source present at the HFF detection level at the transient position. This by itself does not rule out a fainter, variable source at the deeper \emph{Flashlights} level. However, the estimated observed time delay between the transient and its expected counterpart, which should have appeared $\sim0.8$~days earlier, puts an upper limit of $\sim0.35$~days on the rest-frame duration, making a variable source (such as a nova or kilonova) in the source plane unlikely. Note that we cannot entirely rule out an LBV as the source of the transient considering the wide range of features in observed LBV sources. In addition, we cannot rule out the transient being an LBV source in the intracluster medium, as the two observations are separated by 1~yr, but again considering the juxtaposition of the lensed arc and the fact that the transient is detected relatively far from any cluster galaxy makes it a rare event.

A prime candidate for the transient is a lensed star in the source plane. We show that at the transient position microlens distribution with a typical stellar density of $5~{\rm M}_\odot$~pc$^{-2}$ (estimated at the arc position using earlier \emph{HFF} observations) and microlens masses in the mass range [0.08, 1.5]~${\rm M}_\odot$ with Salpeter mass function can lead to microlensing peaks with sufficient magnification to detect O/B-type stars in the source. In addition, the absence of a counterimage can be easily explained if we observed a star while it was crossing a microcaustic. Assuming that the transient is a microlensed star, to further gain insight into the nature of the star we use the F200LP -- F350LP colour. The difference estimate suggests that an O-type star is more preferable than a blue supergiant to explain the observed transient.

If future observations rule out the variable-source hypothesis, then this transient source will be the first example of a lensed star detected with a clear offset from the critical curve location, by about 2--3~kpc at the distance of Abell 370. As shown above, such offsets are not implausible with a sufficient density of microlensing stars within the cluster, consistent with the observed level of the cluster ICL at the transient location. However, because the predicted rate of offset microlensing we find is relatively low, this event may hint at the presence of significant substructure of the dark matter which dominates over the ICL in terms of projected mass density at the transient location. An investigation of substructure from cold-dark-matter subhalos or the pervasive interference from wave dark matter on the de~Broglie scale affecting lensing \citep[e.g.,][]{2020PhRvL.125k1102C} would be warranted with further examples of such large-offset stellar transients identified by our \emph{Flashlights} program, providing the potential to directly constrain the nature of dark matter in the near future.

\section{Acknowledgement}

This research was supported by NASA/{\it HST} grants GO-15936 and GO-16278 from the Space Telescope Science Institute, which is operated by the Association of Universities for Research in Astronomy, Inc., under NASA contract NAS5-26555. 
A.K.M., A.Z., and M.G. acknowledge support by grant 2020750 from the United States-Israel Binational Science Foundation (BSF) and grant 2109066 from the United States National Science Foundation (NSF), and by the Ministry of Science \& Technology, Israel. This work was supported by JSPS KAKENHI grants JP22H01260, JP20H05856, and JP20H00181. A.K.M. would like to thank Lukas Furtak for useful discussions. J.M.D. acknowledges the support of projects PGC2018-101814-B-100 and MDM-2017-0765. 
A.V.F. is grateful for additional financial support from the Christopher R. Redlich Fund and numerous individual donors.
\\
\\
\textit{Facilities}: {\it HST} (ACS-WFC, WFC3-IR)
\\
\\
\textit{Software}: {\texttt{Python}~(\url{https://www.python.org}), \texttt{NumPy}~\citep{harris2020array}, 
\texttt{AstroPy}~\citep{astropy:2018}, \texttt{Matplotlib}~\citep{Hunter:2007}, 
\texttt{DrizzlePac}~\citep{drizzlepac}, \texttt{PythonPhot}~\citep{pythonphot}}

\section{Data Availability}
The data used in this article is publicaly available.

\bibliography{references}
\bibliographystyle{mnras}


\bsp	
\label{lastpage}
\end{document}